 \documentclass[final,5p,times,twocolumn,numeric,sort&compress]{elsarticle}

\usepackage{amssymb}
\usepackage{lipsum}
\usepackage{amsmath}
\usepackage{hyperref}
\usepackage{mathtools}
\usepackage{amsthm}
\usepackage{graphicx}
\usepackage{booktabs}
\usepackage{orcidlink}

\hypersetup{
  colorlinks   = true, 
  urlcolor     = blue, 
  linkcolor    = blue, 
  citecolor   = blue 
}

\journal{Physics of the Dark Universe}
\begin{document}
\begin{frontmatter}



\title{Is cosmological data suggesting a nonminimal coupling between matter and gravity?}

\author[first,second]{Miguel Barroso Varela \orcidlink{0009-0006-9844-7661}\texorpdfstring{\corref{cor1}}{cor1}}
\ead{up201907272@edu.fc.up.pt}

\author[first,second]{Orfeu Bertolami  \orcidlink{0000-0002-7672-0560}}
\ead{orfeu.bertolami@fc.up.pt}
\affiliation[first]{organization={Departamento de Física e Astronomia, Faculdade de Ciências, Universidade do Porto},
            addressline={Rua do Campo Alegre s/n}, 
            city={Porto},
            postcode={4169-007}, 
            country={Portugal}}
\affiliation[second]{organization={Centro de Física das Universidades do Minho e do Porto},
            addressline={Rua do Campo Alegre s/n}, 
            city={Porto},
            postcode={4169-007}, 
            country={Portugal}}
\cortext[cor1]{Corresponding author.}

\begin{abstract}
Theoretical predictions from a modified theory of gravity with a nonminimal coupling between matter and curvature are compared to data from recent cosmological surveys. We use type Ia supernovae data from the Pantheon+ sample and the recent 5-year Dark Energy Survey (DES) data release along with baryon acoustic oscillation measurements from the Dark Energy Spectroscopic Instrument (DESI) and extended Baryon Oscillation Spectroscopic Survey (eBOSS) to constrain the modified model's parameters and to compare its fit quality to the Flat-$\Lambda$CDM model. We find moderate to strong evidence for a preference of the nonminimally coupled theory over the current standard model for all dataset combinations. Although the modified model is shown to be capable of matching early-time observations from the cosmic microwave background and late-time supernovae data, we find that there is still some incoherence with respect to the conclusions drawn from baryon acoustic oscillation observations.
\end{abstract}


\end{frontmatter}


\section{Introduction}
The recent evolution of cosmology to one of the most active fields of research in physics can be attributed to the considerable improvement in the amount and quality of cosmological data captured by increasingly ambitious surveys and collaborations \cite{PrecisionCosmology}. This means that we are in a promising position to confirm or rule out the standard model of cosmology ($\Lambda$CDM), which continues to provide an adequate fit to some sets of data while raising considerable questions about its validity when confronted with some of the main open questions in cosmology \cite{CosmologicalObservations}. The standard model of cosmology is now facing more pressure than ever before, considering the need for dark matter to account for galaxy rotation curves \cite{DarkMatterReview,DarkMatterReview2}, of dark energy to explain the accelerated expansion of the Universe \cite{CosmologicalConstantProblem} and its inability to close the several $\sigma$-wide gap between early and late measurements of the Hubble constant \cite{HubbleTensionReview}. Indeed, a very recent analysis of early James Webb Space Telescope (JWST) subsamples has cross-checked the Hubble Space Telescope (HST) distance ladder and confirmed that this tension cannot be resolved by systematic errors in late-time supernovae data \cite{JWT_HST_Riess2024}. \par
Apart from data provided in the past few years by the Pantheon+ and Sloan Digital Sky Survey (SDSS) collaborations, recent observations from the Dark Energy Survey (DES) and the Dark Energy Spectroscopic Instrument (DESI) data releases have been used to determine the status of beyond-$\Lambda$CDM physics \cite{DES_BeyondLCDM,PantheonCosmologicalConstraints,eBOSS_DR16,DESI}. Although conclusions differ when using different datasets, the global statement is clear - the increasing accuracy of measurements is reducing the standard deviation regions into moderate to severe tension with the standard model of cosmology \cite{DES_DESI_LateTimeDarkEnergy,DES_BeyondLCDM,LCDM_Problem}. Often this is performed by fitting a cosmographic model, which describes the recent history of the Universe in a series expansion of redshift, assuming only the cosmological principle \cite{DES_InverseDistanceLadder,SHOESData}. This is a solid way of extracting key insights into the overall behaviour suggested by the data in terms of the Universe's current expansion rate $H_0$, deceleration $q_0$ and higher-order parameters. Another popular method is to use generalisations of $\Lambda$CDM with the equation of state of dark energy differing from that of a cosmological constant, such as $wCDM$ and $w_0w_aCDM$ \cite{DES_DESI_LateTimeDarkEnergy}, among others. However, all of these methods provide little to no physical explanation of these measurements, instead serving as agnostic red flags pointing to the inadequacy of the standard model. \par

Another approach is to consider particular models with physically motivated modified matter content or alternative theories of gravitation. These include, for instance, the generalised Chaplygin gas \cite{ChaplyginKamenshchik,ChaplyginBilic,Chaplygin,ChaplyginGasVacuum}, Galileon models \cite{GalileonGravity}, $f(T)$ \cite{FTHubbleTension,FTHubbleTension2}, $f(Q)$ \cite{FQHubbleTension}, $f(R)$ \cite{FRHubbleTension,FRHubbleTension2,FRHubbleTension3,FR_Observations} gravity and nonminimally coupled gravity \cite{NMCDarkMatter,NMCAcceleratedExpansion,NMCHubbleTension,ThawingGravity_DESI2024}. Some of these models were tested by the DES Year 5 data analysis of beyond-$\Lambda$CDM models \cite{DES_BeyondLCDM}, and it was found that DES supernovae data combined with eBOSS BAO data suggested that 11 of 15 tested alternative theories were moderately preferred over the standard flat $\Lambda$CDM model.

In this work, we build on the research presented in Ref. \cite{NMCHubbleTension}, where it was shown that a modified theory of gravity with a nonminimal coupling (NMC) of matter and curvature provides a suitable solution to the Hubble tension while at the same time sourcing the late-time acceleration of the Universe's expansion without the need for a cosmological constant \cite{NMCAcceleratedExpansion}. However, that investigation focused on the fundamental aspects of the model's ability to bridge between the early-time cosmic microwave background (CMB) measurements \cite{Planck2018} and data from late-time Universe surveys \cite{SHOESData,PantheonData}. Beyond the Hubble tension, this model has been extensively researched in the context of providing a purely gravitational explanation for dark matter effects in galaxy rotation curves \cite{NMCDarkMatter,NMCDarkMatter2}, constraining the theory's parameters with solar system measurements \cite{NMCSolarSystem,NMCSolarSystem2,NMCSolarSystem3}, modifying the creation of large-scale structure \cite{NMCCosmologicalPerturbations} (this was originally considered in Ref. \cite{NMC_CosmologicalPerturbations_Nesseris} for the simple case $f_2=\lambda R$), altering the propagation of gravitational waves \cite{NMCGravWaves,NMCGWPolarisations} and sourcing cosmological inflation in the early Universe \cite{NMCInflation,NMCInflation2,NMCInflation3}. The aim of the present work is to directly compare the same NMC model to the $\Lambda$CDM predictions by means of a comprehensive statistical analysis of the quality of fit provided to the most recent observational cosmology surveys, which accurately probe different aspects of the large-scale evolution of the Universe over several Megaparsecs (Mpc). This will allow us to quantitatively determine if the presently available data from different collaborations indicates that the NMC model is favoured over the current standard model while also providing accurate predictions for the modified theory's best fit parameters, which we found to be missing from the literature up to this point.  \par
The layout of this paper is as follows. We review the nonminimally coupled model, its associated field equations and their consequences in a Universe described by the Friedmann-Lemaître-Robertson-Walker (FLRW) metric, as well as the numerical method employed to generate cosmological predictions in this theory, in Section \ref{NMCSection}. The different datasets considered in our statistical analysis and their respective characteristics are detailed in Section \ref{DataSection}. We then present the results of each model's best fit to different combinations of datasets and discuss their implications on the standing of $\Lambda$CDM and the NMC theory in Section \ref{ResultSection}. We close the paper in Section \ref{ConclusionSection}, where we draw conclusions and debate possible extensions of our work. We use the $(-,+,+,+)$ signature, define $8\pi G=\kappa^2$ and choose units where $c=1$, although $c$ will be reinstated into equations used to generate predictions which are compared with observational data.

\section{Nonminimally coupled model}\label{NMCSection}
\subsection{Action and field equations}
The nonminimally coupled $f(R)$ model considered in this work was first introduced in Ref. \cite{ExtraForce} and is described by the action
\begin{equation}
    S=\int dx^4 \sqrt{-g} \left[\frac{1}{2\kappa^2}f_1(R)+[1+f_2(R)]\mathcal{L}_m \right],
\end{equation}
where $f_{1,2}(R)$ are arbitrary functions of the scalar curvature $R$, $g$ is the metric determinant and $\mathcal{L}_m$ is the Lagrangian density for matter fields \cite{ExtraForce}. General Relativity is recovered by setting $f_1=R$ and $f_2=0$. The inclusion of a cosmological constant can be achieved by choosing $f_1=R-2\kappa^2\Lambda$ or by including it in the matter content described by $\mathcal{L}_m$. By varying this action with respect to the metric we obtain the modified field equations \cite{ExtraForce}
\begin{equation}\label{FieldEquations}
\begin{aligned}
    (F_1+2 F_2\mathcal{L}_m)G_{\mu\nu}=&(1+f_2)T_{\mu\nu}+\Delta_{\mu\nu}(F_1+2 F_2 \mathcal{L}_m)\\
    &+\frac{1}{2}g_{\mu\nu}(f_1-F_1 R - 2 F_2 R \mathcal{L}_m),
\end{aligned}
\end{equation}
where we have defined $\Delta_{\mu\nu}\equiv\nabla_\mu\nabla_\nu-g_{\mu\nu}\Box$ and $F_i\equiv df_i/dR$. 
By applying the Bianchi identities $\nabla_\mu G^{\mu\nu}=0$ to the field equations above we obtain the modified conservation equation \cite{ExtraForce}
\begin{equation}\label{NonConservationEq}
    \nabla_\mu T^{\mu\nu}=\frac{F_2}{1+f_2}\left(g^{\mu\nu}\mathcal{L}_m-T^{\mu\nu}\right)\nabla_\mu R,
\end{equation}
which is uniquely affected by the nonminimal coupling function $f_2$ and can be simplified to its GR form by reinstating a minimal coupling of matter and curvature, which is achieved by setting $f_2=0$. Throughout this work, we wish to focus solely on the effects of the pure NMC part of the action independently of the minimally coupled part of the $f(R)$ theory, which has been studied in the context of cosmological data in great detail \cite{FRHubbleTension,FRHubbleTension2,FRHubbleTension3}. With this in mind, we set $f_1=R$ and consider $f_2\neq0$. We do not consider a cosmological constant, as the model we will evaluate throughout this work is able to replicate the effects of dark energy and thus source the accelerated expansion of the Universe as a purely gravitational effect of the NMC modified gravity theory \cite{NMCAcceleratedExpansion,NMCHubbleTension}.

\subsection{Cosmological dynamics in nonminimally coupled gravity}
To assess the evolution of the expansion of the Universe we use the flat FLRW metric given by the line element
\begin{equation}
    ds^2=-dt^2+a^2(t)\left(dr^2+r^2 d\Omega^2\right)
\end{equation}
and consider the perfect fluid stress-energy tensor components $T_{00}=\rho$ and $T_{rr}=a^2 p$, where $\rho$ and $p$ are the energy density and the isotropic pressure of the fluid, respectively. Unlike in GR, where only the stress-energy tensor, derived from the matter Lagrangian density, enters the dynamics of the system, the NMC model introduces an explicit dependence on $\mathcal{L}_m$, as seen in Eqs. (\ref{FieldEquations}) and (\ref{ConservationEq}). This turns the choice of the perfect fluid Lagrangian density, formerly being degenerate between $\mathcal{L}_m=-\rho$ and $\mathcal{L}_m=p$, to a non-trivial choice. This has been thoroughly researched in Refs. \cite{LagrangianForm,LagrangianChoice2}, where the physical consequences of each option were investigated and compared. The choice of $\rho$ over $p$ is particularly relevant in the context of late-time modifications of cosmological dynamics in the NMC theory such as the ones considered in this work, as for a matter-dominated Universe one has $p=0$, which would remove the overwhelming majority of the modified theory's effects, given that all modifications are explicitly proportional to the Lagrangian density, unlike in minimally coupled $f(R)$ gravity. A consequence of this is that we would require a significantly larger $f_2(R)$, as this modified model serves as an alternative explanation for the accelerated expansion of the Universe by mimicking dark energy, which makes up the dominant component of the Universe's content at present. Throughout the remainder of this work, we follow the arguments given in those instances of the literature and take the Lagrangian density to be $\mathcal{L}_m=-\rho$. \par
Given the chosen form of the metric and $\mathcal{L}_m$, we can develop the field equations to obtain the modified Friedmann equation \cite{NMCFriedmann}
\begin{equation}\label{ModifiedFriedmann}
    H^2=\frac{1}{6F}\left[2(1+f_2)\tilde\rho-6H\dot F-f_1+F R\right]
\end{equation}
and the modified Raychaudhuri equation \cite{NMCFriedmann}
\begin{equation}\label{ModifiedRaychaudhuri}
    2\dot H+3 H^2=-\frac{1}{2F}\left[2 \ddot F+4H\dot F+f_1-F R+2\kappa^2(1+f_2)p\right],
\end{equation}
where we have defined $\tilde\rho\equiv\kappa^2\rho$ and $F\equiv F_1+2\kappa^2 F_2 \mathcal{L}_m=1-2F_2\tilde\rho$ for simplicity. These are now higher-order equations of the scale factor $a(t)$, as $\dot{F}\sim\dot{R}\sim\ddot{H}$ and $\ddot{F}\sim \dddot H$, and therefore can no longer be solved analytically or with simple computational methods as in GR, where $H$ is directly determined by $\rho$ at all points of the Universe's expansion. The choice of Lagrangian density is such that the conservation equation is unaltered from the minimally coupled theory 
\begin{equation}\label{ConservationEq}
    \dot \rho +3H(\rho+p)=0,
\end{equation}
therefore allowing us to consider the evolution of the energy density of all kinds of matter with respect to the scale factor to be given by the usual expression $\rho\propto a^{-3(1+\omega)}$, which is only dependent on the equation of state parameter $\omega=p/\rho$ ($\omega=0$ for non-relativistic matter, $\omega=1/3$ for radiation).

\subsection{Choice of \texorpdfstring{$f_2(R)$}{f2(R)}}
As we focus on late-time cosmological data, we need to ensure that the effects of the nonminimal coupling are significant in that epoch, thus allowing for sharp conclusions about the theory's parameters. A sensible choice for $f_2(R)$ is one that decouples at high curvatures (early Universe) and becomes increasingly significant at low curvatures (late Universe). This is satisfied by an inverse power law 
\begin{equation}
    f_2(R)=\left(\frac{R_n}{R}\right)^n,
\end{equation}
where $n$ is a positive integer. We can think of this as an isolated term in a putative more complex series expansion of both positive and negative powers of the Ricci scalar which come into play at different characteristic scales $r_c=R_n^{-1/2}$. Multiple terms in this series have been researched in the context of different effects such as dark matter ($f_2\propto R^{-1},R^{-1/3})$ \cite{NMCDarkMatter,NMCDarkMatter2}, dark energy ($f_2\propto R^{-4},R^{-10}$) \cite{NMCAcceleratedExpansion} and inflation ($f_2\propto R,R^3$) \cite{NMCInflation2,NMCInflation3}, among others. \par
As discussed in detail in Ref. \cite{NMCHubbleTension}, this choice of $f_2$ leads to an important assumption about the version of the NMC model considered in this work. As the theory decouples matter and curvature at high redshifts, we can assume that the Universe is governed by GR in the distant past, namely around the CMB epoch. This means that the theory behind early-time measurements, such as those carried out by the Planck experiment, is precisely the same as predicted by the standard $\Lambda$CDM model. Considering that this is the model assumed for the analysis presented in the 2018 Planck results \cite{Planck2018}, which measured several cosmological parameters to an unprecedented degree of precision, we can take those conclusions to determine some of the model's parameters without the need for additional data analysis. Specifically, this constrains the Hubble parameter and matter density at high redshifts to agree with the Planck values extrapolated using the $\Lambda$CDM model, which we denote as $\Omega_m^*=0.315\pm0.007$ and $H_0^*=(67.4\pm0.5)$ km/s/Mpc \cite{Planck2018}. We then write the decoupled initial conditions of the cosmological evolution as \cite{NMCHubbleTension}
\begin{equation}
\begin{array}{lll}
    H^2(z_i)=H_0^{*2}\Omega_m^*(1+z_i)^3 & & R=\tilde{\rho}(z_i)=3H_0^{*2}\Omega_m^*(1+z_i)^3,
\end{array}
\end{equation}
where we have neglected the effect of radiation density, as the initial redshifts we will consider are much lower than the matter-radiation equality redshift $z_{\text{eq}}\sim3000$. The model is then evolved numerically using the modified equations until $z=0$, at which point we can determine \cite{FRHubbleTension}
\begin{equation}
\begin{array}{lll}
   H_0=H(z=0)\neq H_0^* & & \Omega_m=\frac{H_0^{*2}}{H_0^2}\Omega_m^*\neq\Omega_m^*
\end{array}
\end{equation}
as the theory's predicted present parameters. \par
\begin{figure}[ht!]
    \centering
    \includegraphics[width=0.9\linewidth]{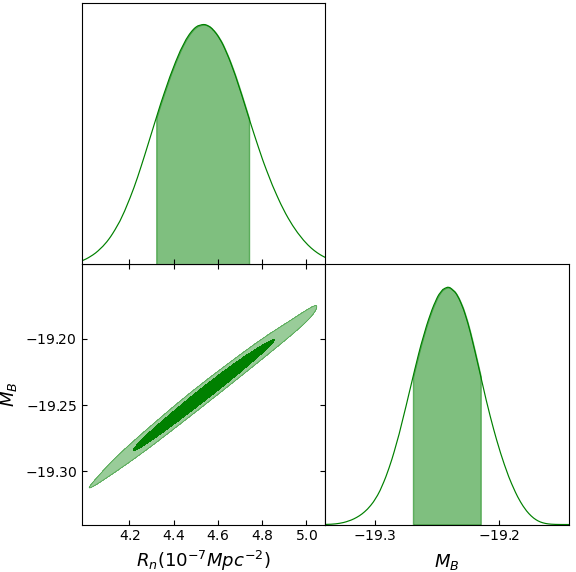}
    \caption{Posterior distributions for the $n=4$ NMC model fit to the Pantheon+SH0ES Cepheid-calibrated distance moduli. The $R_n$ axis is in units of $10^{-7} \ \text{Mpc}^{-2}$ for simplicity. We show the 1$\sigma$ and 2$\sigma$ regions in the 2D posterior and the 1$\sigma$ region in the marginalised 1D posteriors.}
    \label{n4n10PantheonPosterior}
\end{figure}

It is important to note that fixing the initial conditions with Planck data means that we are left with a single free parameter ($R_n$) to be fit from observational data, as we do not vary $n$ directly, instead focusing on fixed positive integer values such as $n=4$ and $n=10$, as done in the latter stages of the analysis described in Ref. \cite{NMCHubbleTension}. This makes the considered NMC models particularly safe from overfitting, especially when comparing their fit quality to that of the flat $\Lambda$CDM model with 2 free parameters ($H_0,\Omega_m$). Nevertheless, when discussing results for the modified theory, we present values for $H_0$ and $\Omega_m$, which are determined directly from the numerical evolution of the model as presented above. With this in mind, these values should not be confused with additional free parameters and are therefore not shown in the posterior distributions in Figure \ref{n4n10PantheonPosterior}.

\subsection{Numerical Method}
The methodology of the numerical evolution of the cosmological parameters is described in detail in Ref. \cite{NMCHubbleTension}. Nevertheless, we present the fundamental aspects of the method employed for the purpose of comparison with observational data. For simplicity, we use $d(\ln a) =H dt$ as our ``time" coordinate. We then evolve our system using the rearranged field equations in the form 
\begin{equation}
    \frac{d^2 \left(F_2 \tilde \rho\right)}{d(\ln a)^2}=-\frac{1}{2}(1-2F_2 \tilde\rho)\left(1-\frac{R}{3 H^2}\right)-\frac{R}{6H^2}\frac{d\left(F_2\tilde\rho\right)}{d(\ln a)}+F_2\tilde\rho\frac{R}{2H^2},
\end{equation}
where we use the $F_2\tilde\rho$ as the dynamic variable instead of the Ricci scalar $R$, which can always be obtained by an inversion of the definition of $F_2(R)$. The Hubble parameter is evolved using its model independent connection with $R$ in the flat FLRW metric
\begin{equation}
    \dot H=\frac{R}{6}-\frac{H^2}{3}\Rightarrow\frac{dH}{d(\ln a)}=\frac{R}{6H}-\frac{H}{3},
\end{equation}
which completes the dynamics of the system, as the energy density is directly calculated from the scale factor, determined simply as $a=e^{\ln{a}}$. \par
The starting conditions for the numerical integrator are obtained from the usual $\Lambda$CDM equations \cite{NMCHubbleTension} with parameters taken from the Planck results due to the decoupling of matter and curvature at high redshifts described above. The precise starting point is adaptively chosen from the theory's parameters to ensure an adequately small value of $F_2\tilde\rho$, here denoted as $\epsilon$. For each value of $n$ and $R_n$ the initial redshift is given by
\begin{equation}
    [F_2 \tilde\rho]_{z=z_i}=\epsilon\Rightarrow z_i=-1+\left(\left(\frac{n}{\epsilon}\right)^{1/n}\frac{R_n}{\tilde\rho_0}\right)^{1/3}, 
\end{equation}
where $\tilde\rho_0=3H_0^{*2}\Omega_m^*$ is the $\Lambda$CDM value of the energy density at $z=0$ taken from the Planck data \cite{Planck2018}.

\section{Observational data}\label{DataSection}
\subsection{Pantheon+SH0ES}
The Pantheon+ dataset\footnote{\url{https://github.com/PantheonPlusSH0ES/DataRelease}} consists of a sample of 1701 cosmologically viable SNIa light curves from 1550 distinct supernovae in the redshift range $0.001<z<2.26$ \cite{PantheonData}. It combines 3 separate mid-$z$ samples ($0.1<z<1.0$), 11 different low-$z$ samples ($z<0.1$) and 4 separate high-$z$ samples ($z>1.0$), each with their own photometric systems and selection functions. This dataset provides values for the distance moduli $\mu$ at different redshifts. Observationally, this is determined from the difference between the corrected apparent magnitude of the SNIa $m_B$ and the absolute magnitude $M_B$, i.e. $\mu_{\text{obs}}(M_B)=m_{B,\text{obs}}-M_B$ \cite{PantheonData}. This quantity is also theoretically defined as 
\begin{equation}
    \mu(z_i)=5\log_{10}\left(\frac{d_L(z_i)}{1\text{Mpc}}\right)+25
\end{equation}
in terms of the distance luminosity 
\begin{equation}
    d_L(z)=(1+z)c\int^z_0\frac{dz'}{H(z')},
\end{equation}
which can be predicted from the analytical or numerical evolution of the Hubble parameter in any particular model. In order to compare predictions with this observational data, we define the log-likelihood and $\chi^2$ values as
\begin{equation}
-2\ln{\mathcal{L}}=\chi^2=\Delta\vec{\mathbf{D}}^T \mathbf{C}^{-1}_{\text{stat+sys}}\Delta\vec{\mathbf{D}},
\end{equation}
where $\Delta\mathbf{D}_i=\mu_i-\mu_{\text{model}}(z_i)$ and we use the covariance matrix provided by the Pantheon+SH0ES data release, which includes both statistical and systematic errors. However, the combination of the form of $d_L(z)$ and the fact that $M_B$ is an additional undetermined parameter of the analysis leads to a degeneracy between $H_0$ and $M_B$ from the theoretical and observational determinations of $\mu$, respectively. This degeneracy can be considerably relaxed by the inclusion of data from SH0ES Cepheid host distance anchors \cite{SHOESData}, which is also provided in the Pantheon+SH0ES data release. We thus modify our analysis in the same way as conducted in Ref. \cite{PantheonData} and now define
\begin{equation}
\Delta \mathbf{D}_i^{\prime}= \begin{cases}\mu_{\text{obs},i}(M_B)-\mu_{i}^{\text {Cepheid }} & i \in \text { Cepheid hosts } \\ \mu_{\text{obs},i}(M_B)-\mu_{\text {model }}\left(z_i,R_n\right) & \text { otherwise },\end{cases}
\end{equation}
where $\mu_i^{\text {Cepheid }}$ is the Cepheid calibrated host-galaxy distance provided by SH0ES and where $\mu_i-\mu_i^{\text {Cepheid }}$ is sensitive to $M_B$ and $H_0$ and is mostly insensitive to other parameters. This modification provides a way of breaking the degeneracy between $M_B$ and $H_0$, as the Cepheid host data is calibrated using the gravitational model-independent properties of Cepheids \cite{R20Cepheids}, thus acting as an anchor to the SNIa data independently of the underlying gravitational theory. This only constrains $M_B$, effectively breaking the degeneracy of the parameters and allowing for an accurate statistical determination of their values when fitting any gravitational theory to the non-calibrated SNIa data. This allows us to constrain both $M_B$ and $R_n$ (which directly determines $H_0$ given the assumptions discussed in Section \ref{NMCSection}) for each NMC model, as well as $\Omega_m$ for the flat $\Lambda$CDM fit used for comparison. For the remainder of this work, we label this dataset as \textbf{PS}.

\subsection{DES-SN5YR}
The Dark Energy Survey (DES) sample of SNIa data\footnote{\url{https://github.com/des-science/DES-SN5YR}} is similar in number to the entirety of the previously discussed Pantheon+ sample, consisting of 1635 photometrically-classified SNIa with redshifts $0.1<z<1.3$ and complemented by 194 low-redshift SNIa (shared with the Pantheon+ sample) in the range $0.025<z<0.1$ \cite{DES_main}. This dataset does not include Cepheid host distance anchors and therefore does not provide a simple procedure to remove the degeneracy between $M_B$ and $H_0$ using observational data. However, as discussed in Ref. \cite{DES_BeyondLCDM}, we can analytically marginalize over the combined parameter $\mathcal{M}=M_B+5\log_{10}(c/H_0)$ when calculating the $\chi^2$ value of a model for this SNIa sample, which would otherwise be calculated in the same manner as for non-Cepheid host SNIa in the Pantheon+ dataset described above but using the covariance matrix provided by the DES Year 5 data release. We thus define the marginalized $\tilde\chi^2$ value as \cite{MB_H0_Marginalisation}
\begin{equation}
    \tilde{\chi}_{SN}^2=\chi_{SN}^2-\frac{B^2}{C}+\ln\left(\frac{C}{2\pi}\right),
\end{equation}
where
\begin{equation}
B=\sum_i(\mathbf{C}^{-1}_{\text{stat+sys}}\Delta\vec{\mathbf{D}})_i
\end{equation}
and
\begin{equation}
C=\sum_i\sum_j\left[\mathbf{C}^{-1}_{\text{stat+sys}}\right]_{ij}
\end{equation}
are defined identically to their original formulation in Ref. \cite{MB_H0_Marginalisation} and their recent application in the DES analysis \cite{DES_main,DES_BeyondLCDM}. While this does not allow us to determine constraints on the value of $H_0$ and $M_B$, it still provides a comprehensive amount of data at mid to high redshifts, which can help constrain the late-time behaviour of the analysed models when combined with baryon acoustic oscillation datasets such as the ones described below. When presenting results, we label this dataset as \textbf{DESYR5}.

\subsection{DESI BAO}
The distance of propagation of baryon acoustic oscillations (BAO) in the primordial fluid in the early Universe is fixed by the sound horizon at the time of baryon decoupling ($z_d\simeq1060$) and left a characteristic imprint in the distribution of matter, which is observable in the late-time galaxy distribution. In their early data release, the Dark Energy Spectroscopic Instrument (DESI) collaboration provides BAO measurements in seven redshift bins from over 6 million extragalactic objects in the range $0.1<z<4.2$ \cite{DESI}. The distance set by BAO serves as a cosmological standard ruler, as their physical scale is set by the sound horizon at the time of baryon decoupling. This can be calculated as
\begin{equation}
    r_d=\int_{z_d}^\infty\frac{c_s(z)}{H(z)}dz,
\end{equation}
where the speed of sound can be determined as
\begin{equation}
    c_s(z)=\frac{c}{\sqrt{3\left(1+\frac{3\rho_B(z)}{4\rho_\gamma(z)}\right)}},
\end{equation}
where $\rho_B$ and $\rho_\gamma$ represent the baryon and photon density, respectively. Note that the integral is evaluated from the baryon decoupling redshift $z_d$ to infinity and is therefore independent of the late-time evolution of the Universe. Particularly for the NMC theory, this means that we can again take the Planck measurement as standing in the NMC theory. We thus use the constraint from Ref. \cite{SoundHorizonPlanck}, where the late-time dependence of the Planck CMB likelihood for $r_d$ was removed, giving $r_d=(147.46\pm0.28)$ Mpc. \par
BAO measurements are generally presented in the form $d_H/r_d$ when considering object pairs that are oriented parallel to the line of sight, where $d_H(z)=c/H(z)$ is the Hubble distance. Conversely, $d_M/r_d$ is presented when they are oriented perpendicularly, with the transverse comoving distance given by
\begin{equation}
    d_M(z)=c\int^z_0\frac{dz'}{H(z')}=\frac{d_L(z)}{1+z}.
\end{equation} 
For redshift bins with low signal-to-noise ratio, the volume-averaged $d_V/r_d$ is given, where
\begin{equation}
    d_V(z)=\left(zd_M(z)^2d_H(z)\right)^{1/3}
\end{equation}
is the dilation scale. For both flat $\Lambda$CDM and the NMC models, we calculate the predicted values of the different distance measurements and compare this with the observational data in an additional $\chi^2$ value. We label this dataset as \textbf{DESI}.

\subsection{eBOSS BAO}
The extended Baryon Oscillation Spectroscopic Survey (eBOSS) is the cosmological survey within the fourth generation of the Sloan Digital Sky Survey (SDSS-IV), which ran from 2014 to 2020. Specifically, the present work uses the available results from the latest data release (DR16) \cite{eBOSS_DR16} combined with a measurement from DR12 \cite{BOSS_DR12}. These earlier results correspond to the two lower redshift bins of the originally published likelihood and are independent of the DR16 results. In total, this dataset provides 8 total measurements, 4 of $d_H/r_d$ and 4 of $d_M/r_d$, along with their corresponding covariance matrix\footnote{\url{https://svn.sdss.org/public/data/eboss/DR16cosmo/tags/v1_0_0/likelihoods/BAO-only/}}, which we analyse as the data from DESI. We label this dataset as \textbf{eBOSS}.

\section{Results and Discussion}\label{ResultSection}
The analysis presented in this work was conducted using the Markov chain Monte Carlo (MCMC) sampler in the \textsc{cobaya} package \cite{COBAYA_paper}, which provides robust code for Bayesian analysis. This package has been used in several recent investigations regarding the statistical treatment of cosmological data, such as the DES beyond-$\Lambda$CDM analysis \cite{DES_BeyondLCDM}, the DESI BAO parameter constraints \cite{DESI} and the preparation of the Euclid mission \cite{EuclidOverview}. The convergence of the MCMC chains was determined by a generalised version of the $R-1$ Gelman-Rubin statistic \cite{GelmanRubin_Rminus1}, which determines the variance in the means of the chains in terms of their covariance. This is the standard convergence criteria used by the \textsc{cobaya} package. Effectively, as $R$ approaches 1, this ensures that all chains are centred around the same point, not deviating from it by a distance that is a significant fraction of the standard deviation of the posterior distribution. In other words, it describes how much sharper the final distribution might become if the analysis was continued indefinitely. Most cases will have confidently converged when $R-1$ is small enough unless dealing with multimodal posteriors \cite{COBAYA_paper}, which is not the case for the analysis conducted in this work, as seen in Figure \ref{n4n10PantheonPosterior}. To ensure the best consistency with the current beyond-$\Lambda$CDM model-testing in the literature, we adopt the same stringent convergence criteria as in Ref. \cite{DES_BeyondLCDM}, where $R-1=0.001$ was taken, in comparison to \textsc{cobaya}'s default criteria of $R-1=0.01$. The priors for all parameters were taken as flat and with adequately broad ranges, thus ensuring that no dependence on these choices was present. \par 
\begin{table*}[ht!]
    \centering
    \begin{tabular}{|c||ccccc|}
        \hline
        \textbf{Model} & $\mathbf{H_0}$ & $\mathbf{\Omega_m}$ & $\mathbf{R_n \ (10^{-7}\textbf{Mpc}^{-2})}$ & $\mathbf{M_B}$ & $\mathbf{\chi^2}$ \\ \hline\hline
        \textbf{PS}&&&&& \\ \cline{1-1}
    {$\Lambda$CDM} & $73.57\pm1.00$    &$0.332\pm0.018$&  \centering-&$-19.246\pm0.029$&1522.99 \\ 
    {NMC $n=4$} & $73.32\pm0.96$    &$0.265\pm0.007$&  $4.54\pm0.21$&$-19.241\pm0.027$&1523.89 \\ 
    {NMC $n=6$} &$72.88\pm0.94$&$0.268\pm0.007$&$4.56\pm0.18$&$-19.273\pm0.028$&1529.68 \\ 
    {NMC $n=10$} & $72.31\pm0.99$    &$0.273\pm0.007$&  $4.79\pm0.18$&$-19.309\pm0.028$&1554.65 \\ 
    \hline\hline
    \textbf{PS+DESI}&&&&& \\ \cline{1-1}
    {$\Lambda$CDM} & $69.99\pm0.66$    &$0.292\pm0.011$&  \centering-&$-19.366\pm0.018$&1561.36 \\ 
    {NMC $n=4$} & $68.21\pm0.57$    &$0.306\pm0.005$&  $3.52\pm0.11$&$-19.388\pm0.019$&1575.92 \\ 
    {NMC $n=6$} &$69.93\pm0.53$&$0.292\pm0.004$&$4.01\pm0.10$&$-19.357\pm0.016$&1559.96 \\ 
    {NMC $n=10$} & $71.45\pm0.55$    &$0.279\pm0.004$&  $4.62\pm0.10$&$-19.333\pm0.016$&1581.60 \\ 
    \hline\hline
    \textbf{PS+eBOSS}&&&&& \\ \cline{1-1}
    {$\Lambda$CDM} & $69.51\pm0.62$    &$0.299\pm0.014$&  \centering-&$-19.378\pm0.017$&1557.25 \\ 
    {NMC $n=4$} & $67.91\pm0.50$    &$0.309\pm0.005$&  $3.47\pm0.09$&$-19.397\pm0.015$&1574.26 \\ 
    {NMC $n=6$} &$69.18\pm0.56$&$0.298\pm0.005$&$3.88\pm0.10$&$-19.379\pm0.017$&1556.86 \\ 
    {NMC $n=10$} & $70.52\pm0.53$    &$0.287\pm0.004$&  $4.47\pm0.09$&$-19.360\pm0.016$&1570.45 \\
    \hline\hline
    \textbf{DESYR5+DESI}&&&&& \\ \cline{1-1}
    {$\Lambda$CDM} & $67.47\pm0.76$    &$0.328\pm0.013$&  \centering-&\centering-&1669.79 \\ 
    {NMC $n=4$} & $66.53\pm0.55$    &$0.322\pm0.005$&  $3.22\pm0.10$&\centering-&1665.32 \\ 
    {NMC $n=6$} &$68.34\pm0.59$&$0.305\pm0.005$&$3.72\pm0.10$&\centering-&1671.92 \\ 
    {NMC $n=10$} & $70.08\pm0.62$    &$0.290\pm0.005$&  $4.39\pm0.11$&\centering-&1716.58 \\ 
    \hline\hline
    \textbf{DESYR5+eBOSS}&&&&& \\ \cline{1-1}
    {$\Lambda$CDM} & $66.75\pm0.66$    &$0.346\pm0.015$&  \centering-&\centering-&1657.62 \\ 
    {NMC $n=4$} & $65.91\pm0.57$    &$0.328\pm0.006$&  $3.11\pm0.10$&\centering-&1657.01 \\
    {NMC $n=6$} &$67.34\pm0.58$&$0.314\pm0.005$&$3.56\pm0.10$&\centering-&1659.44 \\ 
    {NMC $n=10$} & $68.72\pm0.62$    &$0.302\pm0.005$&  $4.15\pm0.11$&\centering-&1694.87 \\ 
    \hline
\end{tabular}
\caption{Best fit parameters determined with different dataset combinations. These results are taken from the \textsc{cobaya} marginalised 1D distributions. The values for $H_0$ and $\Omega_m$ in the NMC model were determined from the fitted values of $R_n$.\label{ParameterTable}}
\end{table*}

Throughout this investigation, we take the $n=4$ and $n=10$ models as standard examples of the inverse power-law NMC theory behaviour, as it was originally found in Ref. \cite{NMCAcceleratedExpansion}. These provided expansion histories that best mimicked a set of chosen deceleration parametrisations based on observational data. This was confirmed in Ref. \cite{NMCHubbleTension}, where once again the same models were shown to be particularly successful in patching the Hubble tension, which was the main goal of that work, while maintaining adequate fits to a select batch of observational data. However, their respective behaviours showed varying capability in matching different cosmological probes, such as cosmic chronometers, BAO measurements and SNIa data. This further reinforces our hypothesis that a full form of $f_2(R)$ may be represented as a sum of different integer powers of $R$ which come into play at independent scales. Although this consideration leads to instabilities in the numerical integration of the equations in the NMC theory, here we additionally consider the $n=6$ theory in an attempt to find a middle ground between the previously considered models.\par
The particular choice of $n=6$ follows from a qualitative analysis of intermediate integer values of $n$, as it best exhibits the late emergence of $n=10$ combined with the smoother variation of $n=4$ \cite{NMCAcceleratedExpansion}. Given that we were searching for a combination of the characteristics of these models, with $n=4$ being favoured by the more extensive SNIa datasets \cite{NMCHubbleTension}, the quality of $n=6$ can be seen as middle ground between 4 and 10 with a slight asymmetry towards 4. Of course, a natural possibility would be to add $n$ as a free parameter, which could be fitted to the data along with $R_n$. However, this would scan through non-integer values of $n$, which is incompatible with the power series form of $f_2(R)$ in powers of $R$, which arises, for instance, from effective field theory arguments. Additionally, as will be discussed in Section \ref{FitQualitySubsection}, one of the upsides of considering this model as an alternative to $\Lambda$CDM is its low number of free parameters, which avoids the risk of overfitting. This provides further motivation for considering only select integer values of $n$, which we do not extend to the full range between 4 and 10 in order to keep computational cost to a minimum. We thus constrain our analysis to the  previously considered values $n=4$ and $n=10$, with the addition of $n=6$ aimed at probing an intermediate behaviour of low and high powers of $R$.

\subsection{Constraints on parameters}
The results for the fitted parameters are shown in Table \ref{ParameterTable}. We consider the combination of each SNIa dataset with each of the individual BAO samples, but do not combine the Pantheon+ and DES Hubble diagram results, as these are built under different assumptions and statistical conditions. Specifically, as described in Section \ref{DataSection}, the PS sample provides Cepheid calibrated results \cite{PantheonData}, which allow for breaking the degeneracy between $M_B$ and $H_0$ (or the associated $R_n$ value in the NMC theory). For this reason, the PS dataset is also considered by itself, while the lack of absolute magnitude calibration for the DESYR5 results forces us to marginalize our posteriors over a combination of $M_B$ and $H_0$ \cite{DES_main} and thus DESYR5 can only be considered in combination with BAO data \cite{DES_InverseDistanceLadder,DES_BeyondLCDM} to be sensibly analysed in the context of the modified theory. For the NMC model, we quote the $H_0$ and $\Omega_m$ values derived from the fitted $R_n$ parameter along with their propagated error, while in $\Lambda$CDM $H_0$ and $\Omega_m$ are the fitted parameters.  \par

\begin{figure}[ht!]
    \centering
    \includegraphics[width=0.9\linewidth]{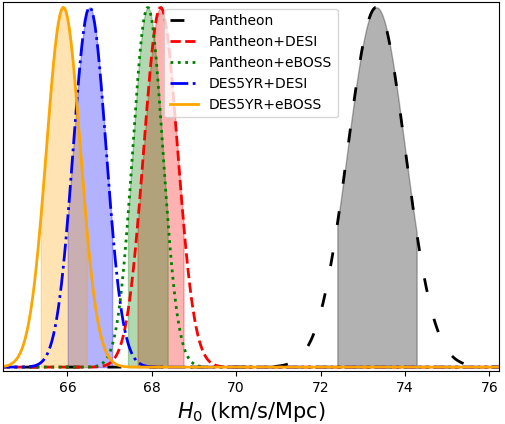}
    \caption{Posterior distributions of $H_0$ for the $n=4$ NMC model fit to different dataset combinations.}
    \label{n4n10_H0_Posteriors}
\end{figure}

The posteriors for the isolated PS data have the highest uncertainties due to the smaller amount of data and the larger number of fitted parameters, as $M_B$ is also fitted for PS SNIa. Nevertheless, they provide solid information on the validity of the statistical analysis built for this work. We find the $\Lambda$CDM results to be in good agreement with their original presentation in Ref. \cite{PantheonCosmologicalConstraints}, serving as a calibration of the \textsc{cobaya} package, which was not used in the original Pantheon+ investigation. Examples of the multi-dimensional posterior distributions for the PS sample in the modified model are shown in Figure \ref{n4n10PantheonPosterior}, where the correlation between $R_n$ (or equivalently $H_0$) and $M_B$ is shown as a narrow diagonal 2D posterior in the $R_n-M_B$ plane, indicating that the Cepheid calibration does not completely remove the correlation between those parameters. All NMC models provide similar values for $H_0$ that are well within error of the model-independent cosmographic result of Ref. \cite{SHOESData}, $H_0=(73.04\pm1.04)$ km/s/Mpc. The absolute magnitude $M_B$ is also within (1-2)$\sigma$ of the $\Lambda$CDM fitted value and particularly in agreement with the SH0ES collaboration value of $M_B=-19.253\pm0.027$ \cite{SHOESData}. The same is no longer true when we combine the PS sample with the DESI and eBOSS datasets. This addition imposes a strong constraint that tends to lower the value of $H_0$ for all considered models, particularly in the case of eBOSS. This decrease in $H_0$ (or equivalently $R_n$) is associated with a decrease in the correlated $M_B$, which becomes significantly lower than for the isolated PS sample while still exhibiting a considerable resemblance between the $\Lambda$CDM and NMC models in terms of their late-time behaviour. This effect of BAO data on the calibration of the absolute magnitude of SNIa has also been found in Refs. \cite{DES_InverseDistanceLadder,MB_BAOData,MB_BAOData2}, where the so-called ``inverse distance ladder" method was applied to calibrate SNIa magnitudes with higher redshift data from BAO. We show the calculated values of $H_0$ in the $n=4$ modified theory for different datasets in Figure \ref{n4n10_H0_Posteriors}. \par

The DESYR5 data is only analysed in combination with each of the BAO measurements from DESI and eBOSS, thus giving comparable uncertainties to the PS+DESI and PS+eBOSS samples. However, the marginalisation of $\mathcal{M}(H_0,M_B)$ in the calculation of the DES SNIa $\chi^2$ value means that the determination of $H_0$ follows from the respective BAO data, which already for the PS sample had a tendency to lower the fitted value of $H_0$, thus leading to even lower values of $H_0$ for both $\Lambda$CDM and NMC models. For all samples that include BAO data, we find that the $H_0$ value in the NMC model grows with the exponent $n$, meaning that for large enough $n$ one could always find a larger $H_0$ at the cost of a typically worse fit quality. A similar result was found for a more limited amount of data in Ref. \cite{NMCHubbleTension}, where it was determined that considering BAO data led to considerably smaller values of $H_0$ in the NMC model, similarly to what happens in $\Lambda$CDM. This work confirms that original tendency with more cohesive choices of data from eBOSS and DESI, which allow us to consider the correlation between different data points unlike when using scattered results from the literature. Even with this new analysis, we find that the NMC model can still only cohesively patch the Hubble tension when considering cosmological model-independent data from SNIa, such as that from the Pantheon+ sample. As discussed in Section \ref{DataSection}, the BAO measurement method finds values for $d_i/r_D$, which we then compare to theoretically predicted values of $d_i$ along with a sound horizon value based on $\Lambda$CDM Planck data, which we use regardless of the late-time deviation from the standard model in the modified theory as the chosen form of $f_2(R)$ corresponds to a minimal coupling at high redshifts. We are thus led to the conclusion that this pending ``BAO tension" can only be resolved in the NMC model by the addition of early-time modifications ($f_2\propto R^n$, $n>0$) or the consideration of systematic errors in the determination of observational BAO data. Concerning the former, this is seen as a remote possibility, considering the extreme precision with which Planck results are calculated in agreement with $\Lambda$CDM \cite{Planck2018}.

\subsection{Fit quality}\label{FitQualitySubsection}
For each model and dataset, we present the calculated values for $\chi^2$ and two criteria were selected to compare the quality of fit of each model. The first of these is the Akaike Information Criterion (AIC), defined by 
\begin{equation}
    AIC=2k-2\ln\mathcal{L}^{max},
\end{equation} 
where $k$ is the number of fitted parameters in the model and $\mathcal{L}^{max}$ is the maximum posterior likelihood determined from the MCMC analysis \cite{AIC}. Following the criteria suggested in Ref. \cite{AIC_Evidence}, it is established that $\Delta AIC>2$, 5 and 10 respectively indicate weak, moderate, and strong evidence against the model with the higher $AIC$ value. This criterion sets a linear penalty for models with greater numbers of parameters. The second of these measures is the Bayesian Information Criterion (BIC), which is defined by
\begin{equation}
    BIC=k\ln N-2\ln\mathcal{L}^{max},
\end{equation}
where $N$ is the number of data points in the analysed sample \cite{BIC}. This criterion has a similar penalty to that of the $AIC$ for small samples, but penalises models with more fitted parameters particularly harshly for large amounts of data points, as is the case for the SNIa datasets considered in this work. We follow the same comparison method for $\Delta BIC$ as the one previously described for $\Delta AIC$ \cite{AIC_Evidence}. Considering the large number of data points used in the fitting process, where $N\sim\mathcal{O}(10^3)$, the $BIC$ can serve as a solid alternative to the $AIC$, prioritising models with small numbers of fitted parameters, rewarding simplicity in the explanation of data and thus avoiding over-fitting. These results are presented in Table \ref{AIC_BIC_Table}. \par

\begin{table}[ht!]
    \centering
    \begin{tabular}{|c||ccc|}
        \hline
        \textbf{Model} & $\mathbf{\chi^2}$ & $\mathbf{\Delta AIC}$ & $\mathbf{\Delta BIC}$ \\ \hline\hline
        \textbf{PS}&&& \\ \cline{1-1}
        $\Lambda$CDM & 1522.99 &0.0&  0.0 \\ 
        NMC $n=4$ & 1523.89 &-1.10&  -6.54\\ 
        {NMC $n=6$} &1529.68&4.69&-1.25 \\ 
        {NMC $n=10$} & 1554.65 &19.76&  14.32\\ 
        \hline\hline
        \textbf{PS+DESI}&&& \\ \cline{1-1}
        $\Lambda$CDM & 1561.36 &0.0&  0.0 \\ 
        NMC $n=4$ & 1575.92 &12.56&  7.12\\ 
        {NMC $n=6$} &1559.96&-3.40&-8.84 \\ 
        {NMC $n=10$} & 1581.60 &18.24&  12.80\\ 
        \hline\hline
        \textbf{PS+eBOSS}&&& \\ \cline{1-1}
        $\Lambda$CDM & 1557.25 &0.0&  0.0 \\ 
        NMC $n=4$ & 1575.92 &16.67&  11.23\\ 
        {NMC $n=6$} &1556.86&-2.39&-7.83 \\ 
        {NMC $n=10$} & 1581.60 &22.35&  16.91\\ 
        \hline\hline
        \textbf{DESYR5+DESI}&&& \\ \cline{1-1}
        $\Lambda$CDM & 1669.79 &0.0&  0.0 \\ 
        NMC $n=4$ & 1665.32 &-6.67&  -12.11\\ 
        {NMC $n=6$} &1671.92&0.13&-5.31 \\ 
        {NMC $n=10$} & 1716.58 &14.79&  9.35\\ 
        \hline\hline
        \textbf{DESYR5+eBOSS}&&& \\ \cline{1-1}
        $\Lambda$CDM & 1657.62 &0.0&  0.0 \\ 
        NMC $n=4$ & 1657.01 &-2.61&  -8.05\\ 
        {NMC $n=6$} &1659.44&-0.18&-5.62 \\ 
        {NMC $n=10$} & 1694.87 &35.25&  29.81\\ 
        \hline
    \end{tabular}
    \caption{Fit quality of models to the different datasets. We show the $\chi^2$ values along with the AIC and BIC measures, which are given in terms of their difference to those of the $\Lambda$CDM model. \label{AIC_BIC_Table}}
\end{table}

The $n=10$ model provides the worst overall fit to all datasets, being strongly disfavoured in comparison to $\Lambda$CDM by both the AIC and BIC measures, with $\Delta BIC$ and $\Delta AIC$ both being greater than 10 for almost all data combinations. This follows from the large value of the exponent in $f_2(R)$, which delays the growth of the NMC effects until considerably late in the Universe's expansion and thus creates a sharp variation in the form of $H(z)$ which is not easily compatible with SNIa data, as initially seen in Ref. \cite{NMCHubbleTension}. Comparatively, the $n=4$ model has significantly varying degrees of success in comparison to $\Lambda$CDM, ranging from strongly favoured to strongly disfavoured in the $BIC$ values. We find this version of the NMC theory to be better adapted to combinations of DESYR5 and DESI/eBOSS data, which likely follows from the marginalisation of $M_B$ and $H_0$ in the DES analysis. It is worth noting that this model is favoured over $\Lambda$CDM when considering the isolated Pantheon+ sample, with the exact opposite being true when we combine this with BAO data. This shows how the $n=4$ theory can provide a solid explanation for model-independent SNIA observations while being unable to simultaneously provide a good fit to BAO measurements. \par

As pointed out above, the $n=6$ model was investigated in an attempt to find a middle ground between the smoother transition of $n=4$ and the late-time sharp behaviour of $n=10$, with the goal of bridging between SNIA and BAO observations. This model's fit to the PS sample is weakly disfavoured to the $\Lambda$CDM model when considering their respective $AIC$ values ($\Delta AIC=4.69$) and leads to no conclusive preference from the $BIC$ measure, which penalises the additional fitted parameter in $\Lambda$CDM. It underperforms $n=4$, as expected from the tendency for $n>4$ to have a sharper transition at late times. However, the conclusions are much clearer when also considering BAO data. Unlike the case of $n=4$, the BAO samples improve this NMC model's fit. In fact, the $\chi^2$ value is lower to the point of overperforming $\Lambda$CDM in pretty much all of the corresponding $AIC$ and $BIC$ values, with our results for $BIC$ indicating moderate evidence for preference of the $n=6$ NMC model over $\Lambda$CDM.

\section{Conclusions}\label{ConclusionSection}
In this work, we have compared a modified theory of gravity with nonminimal coupling between matter and curvature to various recent data from cosmological surveys. To do this, we have reviewed the theoretical consequences of the NMC model on the Universe's dynamics at large scales, particularly at late times. We then discussed the different datasets available and detailed the statistical analysis conducted along with the parameters chosen to quantify the fit quality of each model. \par
We have presented the best fit parameters for 3 variations of the NMC theory, two of which were chosen from previous research on the model's ability to patch the Hubble tension ($n=4,10$) \cite{NMCHubbleTension}, with the other ($n=6$) being tested in an attempt to harness the best qualities of the late-time behaviour of each of the other two \cite{NMCAcceleratedExpansion}. We found that two of these models fit the Cepheid-calibrated Pantheon+ SNIa sample at a level superior to $\Lambda$CDM, predicting $H_0$ values within error of the standard model and the model-independent cosmographic approach from the SH0ES collaboration \cite{SHOESData}, thus effectively fixing the tension between conclusions drawn from supernovae and CMB data due to the model's CMB-based initial conditions and late-time deviations from $\Lambda$CDM. This is particularly non-trivial, as this theory does not need a cosmological constant to match SNIa data, meaning that the NMC model removes the cosmological constant from the problem. However, when combining data from supernovae with BAO measurements, we find that both the standard model and the modified theory tend to give lower values of $H_0$ which are several $\sigma$ away from the ones obtained for the isolated supernovae dataset. This means that although the NMC theory bridges the gap between early-time CMB and late-time SNIa observations, which is not possible in $\Lambda$CDM, it is still not able to provide a consistent explanation for the discrepancy between the SNIa+CMB and SNIa+CMB+BAO dataset conclusions on cosmological parameters.  \par
We then tested the fit quality of each model through two distinct statistical criteria, $AIC$ and $BIC$, which assign different penalties for larger numbers of fitted model parameters and allow us to quantitatively compare the modified theory to $\Lambda$CDM. We found that the $n=10$ model consistently fits SNIa+BAO data worse than all other models, ruling it out with strong evidence from both $AIC$ and $BIC$. The $n=4$ model was moderately to strongly preferred over $\Lambda$CDM for 3 of the 5 datasets, struggling only with the combination of Pantheon+ and eBOSS/DESI BAO data, which follows from the tendency for BAO observations to give $H_0\sim69$ km/s/Mpc, significantly below the Pantheon prediction of $H_0\sim73$ km/s/Mpc. The $n=6$ model is consistently preferred over $\Lambda$CDM with moderate to strong evidence, as it can satisfy both the low-$z$ SNIa points and the high-$z$ BAO data. We thus conclude that current cosmological data suggests the presence of a nonminimal coupling of matter and curvature over that of the minimally coupled standard theory. \par
This work's results raise interest in several possible extensions to its research. Firstly, the continuous release of new and improved data from cosmological surveys allows the NMC model to be put to the test against $\Lambda$CDM and other alternative theories with an increasing degree of accuracy, meaning we can draw more assertive conclusions about evidence for or against physics beyond the standard model of cosmology. It would be particularly relevant to probe the ``BAO tension" which is still present in the NMC model, its connection to assumptions about physics at the CMB epoch and possible solutions provided by the NMC model to explain the standing difference between BAO measurements and CMB+SNIa data, which are compatible in the modified theory. A further issue to be addressed is the $\sigma_8$ tension \cite{Planck_Lensing_sigma8,sigma8_2,Sigma8}, whose analysis involves different datasets and concerns the dynamics of cosmological perturbations in the NMC theory \cite{NMCCosmologicalPerturbations} and is thus left as the topic of a future investigation.

\section*{Acknowledgments}
    The authors would like to thank Alex Bernardini and Rogério Rosenfeld for the discussions that motivated this investigation. The work of one of us (O.B.) is partially supported by FCT (Fundação para a Ciência e Tecnologia, Portugal) through the project 2024.00252.CERN.

\bibliographystyle{elsarticle-num-names} 
\bibliography{ReferencesRevised}


\end{document}